# Stealth-MITM DoS Attacks on Secure Channels

Amir Herzberg and Haya Shulman

*Bar-Ilan University, Ramat-Gan, 52900, Israel,*
{amir.herzberg,haya.shulman}@gmail.com

**Abstract.** We define *stealth Man-in-the-Middle* adversaries, and analyse their ability to launch *denial and degradation of service (DoS)* attacks on secure channels. We show realistic attacks, disrupting TCP communication over secure VPNs using IPsec. We present:

1. First amplifying DoS attack on IPsec, when deployed without anti-replay window.
2. First amplifying attack on IPsec, when deployed with a 'small' anti-replay window, and analysis of 'sufficient' window size.
3. First amplifying attack on IPsec, when deployed with 'sufficient' window size. This attack (as the previous) is realistic: attacker needs only to duplicate and speed-up few packets.

We also suggest a solution designed to prevent the presented attacks, and to provide secure channel immune to degradation and other DoS attacks. Our solution involves changes (only) to the two gateway machines running IPsec.

In addition to their practical importance, our results also raise the challenge of formally defining secure channels immune to DoS and degradation attacks, and providing provably-secure implementations.

## 1 Introduction

Denial/Degradation of service (DoS) attacks pose an ever growing threat to Internet services and applications. Secure channel protocols, with IPsec [23, 34] being the predominant one, are used to securely connect virtual private networks (VPN), i.e., authenticate data and origin, ensure confidentiality, and prevent replay. IPsec is designed to protect against MITM adversaries that can eavesdrop on the communication and inject spoofed segments into the message stream. It is widely believed, and also specified e.g., in [23], that IPsec also defends higher-layer traffic from DoS attacks when attacker has limited resources (e.g., can only block, inject or reorder a limited number of packets). Defense against DoS attacks is often an important consideration in adopting IPsec for protecting a VPN (rather than say using SSL/TLS [21, 12]). We show that this belief is not precise by presenting several DoS attacks on TCP when used over IPsec.

Our attacks raise the following question: what are the properties that secure channel should satisfy to protect against performance degradation attacks? Existing works do not analyse the properties that secure channel protocols should possess to protect against denial of service attacks. There are works that attempt to define what a secure channel is, e.g., [11], but they fail to capture performance analysis of secure channel, i.e., efficiency and resistance to denial of service attacks. Herzberg and Yoffe [20] present a framework that allows to define specifications that capture such properties, and suggest further research on defining secure channel protocols within that framework. However, they do not present such specifications for DoS-preventing secure channel protocols, or demonstrate that existing secure channel protocols fail to protect against DoS. Our work provides such demonstration; we hope that it will prompt research leading to such specifications and provably-secure DoS-preventing channels. Specifically, we show that although IPsec provides an anti-replay mechanism that is targeted at protecting against denial of service by detecting and discarding spoofed packets injected by a MITM adversary, it fails to counter denial of service attacks. We show denial of service attacks that exploit congestion control mechanism of TCP. In each section we present different techniques for exploiting the vulnerabilities of TCP congestion control mechanism, which rely on slightly different adversarial model. We show attacks generating forward path reordering, resulting in reduced transmission rates and increased number of retransmissions, and attacks resulting in reverse-path reordering which lead to bursty transmissions and eventually to increased network congestion. We then investigate the impact that these attacks can have on TCP performance (when run over IPsec), and identify and describe the vulnerabilities that can be exploited to defeat TCP congestion control mechanism. The attacks we present rely on standard behaviour of correctly implemented TCP congestion control mechanism. We analyse worst case scenarios and our results should be regarded as upper bounds on the damage that weakest possible attacker can inflict. In addition, we demonstrate the necessity for and motivate the anti-replay mechanism of IPsec, by presenting simple attacks on TCP congestion control mechanism when IPsec is used without the anti-replay window. We also investigate the correct *size* of IPsec's anti-replay window, and show attacks when incorrect window size is used. To best of our knowledge, IPsec's anti-reply mechanism was not analysed in malicious setting prior to this work. We show successful reduction of quality (RoQ) attacks on TCP by stealth MITM adversary even when infinitely large IPsec anti-replay window is employed.

In Section 5.3 discuss solutions to combat the reordering attacks (whether malicious by adversary, or benign, due to network congestion), and present a reordering tolerant tunneling protocol in IPsec gateway, to address the network reordering. The solution can be either in TCP in every host or in IPsec firewall. We suggest a mechanism in firewalls, which requires minimal changes to existing implementations, to combat the attacks presented in the rest of this paper. Our goal is not to require changes in the TCP protocol in every host separately, but to apply the modification to the firewall, and as a result to protect subnet of hosts. Many private networks connected to the Internet are protected by firewalls. Firewall protection is based on the idea that all packets destined to hosts behind a firewall have to be examined by the firewall. Our solution is comprised of two phases: first detection of an attack, then prevention of an attack. The main idea of our proposition is to delay duplicate packets the duplicate congestion response.

Similar attacks can be applied to other protocols, e.g., to the widely used tunneling Generic Routing Encapsulation (GRE) mechanism, see [13, 14]. According to [13], GRE specifies a protocol for encapsulation of an arbitrary protocol over another arbitrary network layer protocol, and is a common way to achieve tunneling of IP encapsulated inside IP. GRE does not provide authentication, i.e., it is vulnerable to spoofing adversary; to perform denial of service against GRE, an attacker can simply send a segment with a higher sequence number. To prevent this type of attacks, it is suggested to run GRE over IPsec, however, as we show in this work, IPsec does not protect against this type of attacks.

In all our attacks we assume a stealth MITM attacker model, presented in Section 2.2, that can with minimal effort significantly degrade the performance of communication over TCP. Our attacker may be restricted in its eavesdropping capability (may be able to eavesdrop on one network segment but not the other), as well as in the number of (spoofed) packets that it can inject. For instance, in wireless network attacker can only eavesdrop on wireless communication, and may be able to inject segments in the wired access network. Often attackers may be limited in their spoofing ability, e.g., attacker is able to disrupt communication by infiltrating a small device which has a limited power. In addition, attackers typically prefer to avoid detection, thus spoofing a limited number of segments. Note that our attacks exploit the congestion control of TCP, by injecting duplicate segments. This strategy allows attacker to evade DoS detection mechanisms, e.g., consider a sequence of routers on the path between source and destination, where the attacker controls one of the routers. The router simply duplicates some of the segments that traverse it, and reroute them via an alternative path. Thus the malicious router cannot be traced back. On the other hand, if the router simply dropped occasional segments, this could be detected, and the attacker would be traced back to the malicious router. For more details on attacks on wireless networks by MITM adversary (and limitations) can be found in [30]. Similar attacker model was considered in [33], which considers an Explicit Congestion Notification (ECN) with IPsec. We discuss this briefly in Related Works in Section 1.1.

### 1.1 Related Works

**Explicit Congestion Notification (ECN)** There is one reference (which we are aware of) to DoS over IPsec, w.r.t. Explicit Congestion Notification (ECN), in [33]. In [33], authors mention a vulnerability of IPsec to denial of service when using ECN. If the IPsec gateway at the exit of the tunnel does not copy the ECN bit, then it ruins the ECN mechanism; on the other hand, if the gateway copies the ECN bit, then an attacker can degrade performance. The attack can be launched since the authentication that IPsec performs does not protect the ECN bit. However, there is noanalysis of this attack; such analysis is rather similar to the analysis we present, of similar attacks. In addition, our attacks work even if ECN bit is not used, as well as if the recommendation of the RFC not to copy the ECN bit from tunneled packets is followed. Note, that the authors of [33] consider similar adversarial model to ours, i.e., they consider a 'weaker MITM' attacker model like the one we present and define in Section 2.2, although we also consider duplications, and do not consider modifications to legitimate packets, e.g., turning on/off ECN bit.

**Denial/Degradation-of-Service (DoS) Attacks** Denial/Degradation of Service (DoS) attacks, and especially Distributed DoS (DDoS) attacks, pose a serious threat to Internet applications. In the last years, DoS attack methods and tools are becoming more sophisticated, effective, and also more difficult to trace to the real attackers. TCP targeted (low-rate) *Shrew* attacks, [26] exploit the retransmission timeout (RTO) of TCP, by transmitting short traffic pulses of RTT scale length, of low average volume of RTO scale periods, causing TCP flows to continually timeout. The result is near zero TCP throughput. Due to the nature of the attack traffic it can be hard to distinguish it from other legitimate traffic, e.g., video. Low-rate TCP attacks are much harder to detect, and require much weaker attacker capabilities, i.e., the attacker can simply generate bursty UDP flows of low average rate.

Low-rate TCP targeted *Reduction of Quality (RoQ)* attacks are another type of low-rate TCP attack, introduced in [16, 17, 29], where attacker exploits the TCP AIMD mechanism causing TCP performance degradation. The main difference is that RoQ attacks do not require precise timing (to tune to the RTO frequency). The RoQ attacks are even more difficult to detect and block, since they do not operate at specific intervals. In [29] authors suggest a type of attacks similar to RoQ attacks, i.e., the pulsing attacks, which are targeted at TCP applications. The pulsing attacks can be categorised into two models: *timeout-based* attacks, and *AIMD-based* attacks, depending on the timing of the attack pulses w.r.t. congestion window of TCP. During the attack, pulses of malicious traffic are sent to a victim, resulting in packet losses. Authors of [29] show that even a small number of attack pulses can cause significant throughput degradation.

Recently, in [1], a new denial of service attacks, dubbed *JellyFish*, were exhibited. JellyFish attacks target TCP congestion control mechanism of TCP flows, by having the relay nodes misorder, delay or drop packets which they are expected to forward.

Low rate TCP targeted attacks can be prevented by using secure channel protocol between the gateways, e.g., LOT in [19], and using mechanisms that provide quality of service by differentiating traffic, e.g., DiffServ [6]. Namely, when employing DiffServ, flows are given different priority, and flows over a secure channel can be given higher priority, and will be reserved space in routers buffers. Alternately, non-conforming packets can be dropped or given a lower priority and placed in different queues. IPsec anti-replay window prevents replays of the communication exchanged by the legitimate parties by discarding duplicate segments at the receiver. A large anti-replay window can require significant memory resources. There are optimisation works on anti-replay mechanism, e.g., [22, 15, 39]. The justification for an anti-replay mechanism is a MITM adversary which can mount a DoS attack by replaying messages exchanged by the legitimate parties. In a replay attack an adversary sends a copy of previously transmitted, legitimate message between a sender and a receiver, see [35] for more details. When the replayed packet reaches the destination, it will be passed to the transport layer buffer. Duplicate messages degrade performance and is an obvious motivation for anti-replay window. According to [23, 22, 15, 39], the anti-replay mechanism of IPsec is used to secure IP against an adversary that can insert possibly replayed messages in the message stream, and as a result prevent denial of service attacks. In particular, the authors of [39] present *robustness* to DoS attacks as one of the requirements from anti-replay mechanism, and claim that the possibility that packets will be dropped is traded with the prevention of replay attack. We show that even large enough anti-replay window cannot prevent DoS attacks, and in Section 5 we present a new type of *low-rate* TCP attacks, the *speedup attacks*, which significantly degrades performance even when infinitely large IPsec window is used. Our attacks hold also against the suggested proposals for improvements to the anti-replay mechanism of IPsec. In particular, the improved anti-replay window scheme of [39], the control shift mechanism of [22], and the two windows suggested by [15].

### 1.2 Contributions

The contributions of this work can be summarised as follows:

- We identify an important attack model, the stealth MITM attack, which was not explicitly defined prior to this work.
- Justification and analysis of IPsec anti-replay mechanism, Section 3. We show how to compute optimal IPsec anti-replay window to prevent packets loss due to speed up attacks in Section 4.
- *Low-rate* attacks on TCP when running over IPsec, which work even when a large IPsec anti-replay window is employed, Section 5. We analyse our results with a simple analytical model of TCP performance degradation.
- We propose a solution, the reordering tolerant tunneling protocol (RTTP), Section 5.3, in IPsec gateways, to prevent the stealth-MITM attacks.
- Conceptual contribution: we initiate investigation of the performance properties that secure channel protocols should provide.

## 2 Model

In this section we present the scenario which we consider in the paper, we model and motivate the attacker, and give assumptions on the communication.

### 2.1 Scenario and Attack Model

Consider the scenario presented in Figure 1(a), with a virtual private network between two branches, both connected via gateways $GW1, GW2$ to the Internet. All the communication between the branches is over IPsec, using IPsec's ESP mode with authentication. For simplicity, we assume that the clients are located behind $GW1$, and the servers are located behind $GW2$. The clients send requests to download files from servers, and the servers send the requested files in response; upon each correctly received data segment, a client generates and send an acknowledgment (ACK). We assume that all communication is over TCP. We consider an attacker located on the Internet between the two gateways, $GW1$ and $GW2$, in Figure 1(a). The attacker is able to eavesdrop on the communication and inject (a limited number of) packets into the message stream, but cannot drop legitimate packets. Attacker that drops packets can easily disrupt communication and mount a denial of service attack. However, in reality attackers often do not have such abilities. Even when attackers can drop packets they often prefer to refrain when an alternative exists, in order to avoid detection. In addition, we show that the damage inflicted by our stealth attacker is more severe than a damage that a MITM attacker can mount by dropping a limited number of network packets. Note that attacker blocking all communication can clearly mount a DoS, yet we are interested in sophisticated *amplification* attacks where attacker spends considerably less resources w.r.t. the resulting damage. More details on the attacker are presented in the next subsection.

### 2.2 Stealth-MITM Adversary Models

In this work, we define and consider a weak duplicating MITM adversary model, the stealth adversary that cannot delay or drop packets; the 'classical' man-in-the-middle (MITM) adversary can eavesdrop, intercept communication and inject spoofed packets into the message stream. Furthermore, like in low rate attacks [26], we restrict the attacker's ability to send (inject) spoofed and/or duplicated packets. Specifically, we believe a realistic model would be to define a quantified, $(\rho, \sigma)$-limited stealth MITM adversary following the 'leaky bucket' approach. Namely, an $(\rho, \sigma)$-limited MITM adversary is one who can send, during any interval of length $T$, at most $\rho \cdot T + \sigma$ spoofed and/or duplicated packets. These limitations are weaker compared to those of low-rate attacks, e.g., [26, 17, 29], since the attacker is not just limited in the amortised traffic, but also cannot create bursts of traffic. In particular, the bursts are limited by $\sigma$, i.e., an $(\rho, \sigma)$-limited

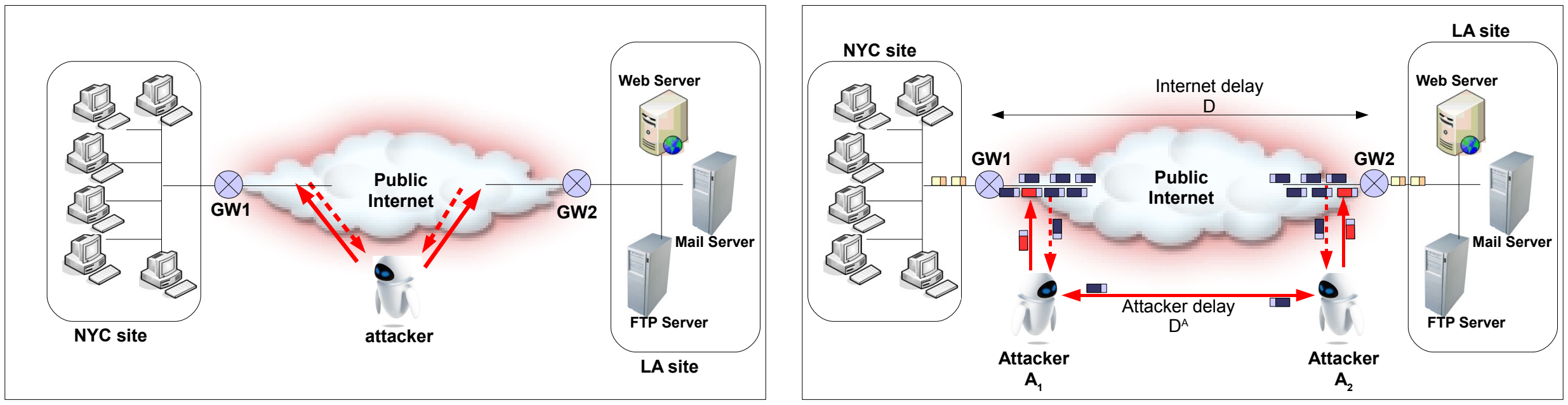

(a) MITM adversary attacking a Virtual Private Network (VPN) between two sites.

(b) MITM adversary between two Virtual Private Networks (VPN) with multiple sites.

**Fig. 1.** Virtual private network behind gateway $GW1$ with users accessing a web server located behind $GW2$. IPsec is used for protection, and a MITM adversary is located on the Internet. Figure 1(b) is an abstract presentation of an attacker that can both eavesdrop on communication in one site and inject spoofed segments in another.

attacker can create a $\sigma-$burst which is a burst of $\sigma$ segments. We show that even this weaker attacker can dramatically degrade performance, even when the communication is protected by IPsec.

We consider (weak) MITM attackers, this implies that the packets they inject can depend on the packets they eavesdrop. In fact, since the communication between the two gateways is authenticated (using IPsec), it follows that the adversary can effectively only duplicate packets, and possibly 'speed up' delivery of a duplicate so it will arrive before the regularly-sent packet, e.g., via an alternative path. Note that the attacker may be limited in the direction in which it can inject spoofed segments, e.g., can only duplicate segments sent from NYC site to LA site in Figure 1(a), but cannot duplicate segments in the other direction.

In each attack we present we use a slightly different variant of the $(\rho, \sigma)$-limited MITM attacker; the different variants of the attacker model are illustrated in Figure 2 and defined below; The weakest stealth MITM adversary (Figure 2(a)) can duplicate packets. A stealth MITM attacker in Figure 2(b), can also speed up packets[1] via faster route, in addition to its ability to duplicate packets. In Figure 2(c), the adversary can duplicate and speed up multiple (e.g., three) packets, not just one. The justification of our adversarial model, is that we focus on the use of IPsec, and IPsec is necessary only when there is concern about MITM. In particular, the anti-replay mechanism that IPsec employs is used to prevent injection of duplicate segments, by identifying and discarding replayed packets. This type of attack can be performed by attacker that can eavesdrop and inject spoofed packets, i.e., a MITM attacker.

As we mentioned before, we are not interested in trivial 'flooding' attacks where the attacker achieves degradation by spending resources proportional to the performance degradation achieved, e.g., attacker injected two packets thus the link carries additional load, and IPsec has to inspect two more packets, resulting in some degradation performance but also attacker's 'cost' is proportional. We are focus on *amplification* attacks where the attacker pays minimal resources with respect to the inflicted damage, e.g., injects three packets, but with a devastating result on the attacked flows.

### 2.3 Communication Assumptions and Model

We assume that the segments arrive with fixed latency which is known to the attacker.In addition, we assume that the attacker (similarly to other network entities) is subject to some non-zero network delay, which may be smaller than that of the legitimate parties, and which the attacker cannot change. Throughout the paper we denote by RTT (Round Trip Time) the time it takes to transmit a segment from a client into the network

[1] Further research should be conducted to consider the damage that attackers without speed-up capabilities, i.e., with the same delay (as the legitimate communicating parties), can inflict when using an alternative route than the one which the legitimate packets traverse.

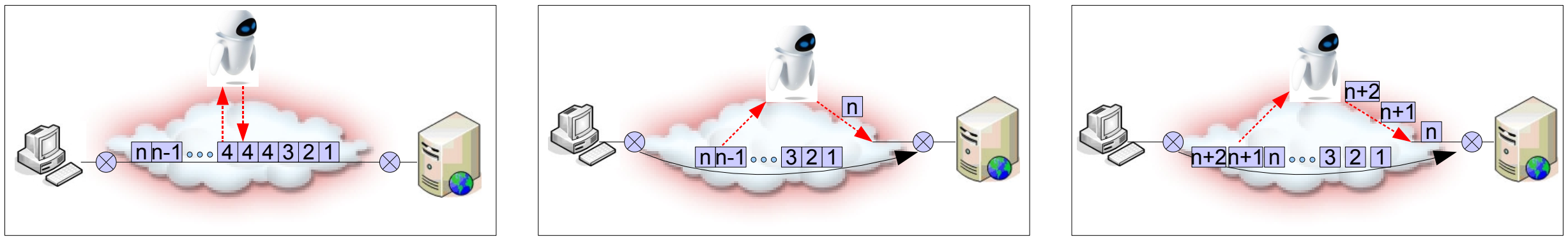

(a) Duplicating stealth-MITM adversary.

(b) Speeding-up stealth-MITM adversary.

(c) Multiple packets speeding-up stealth-MITM adversary.

**Fig. 2.** Duplicating Stealth MITM Adversary Models.

and to receive an acknowledgment (ACK) for it in response. Namely, the RTT is the sum of transmission time of the segment, the propagation delay, and the transmission of ACK and its propagation delay back to the sender, including any queuing and processing delays involved.

The attacks we present apply to standard TCP implementations [32]; TCP state machine is in Figure 3 (from [25]). We assume that the connection is always open, and that the sender sends full sized segments as fast as its congestion window allows. For simplicity (only), assume that the recipient acknowledges every segment received from the server, i.e., no delayed ACKs[2]. Also for ease of exposition, we work with segments instead of bytes (which is what TCP actually sends).

Let $cwnd(t)$ be the congestion window size at time $t$. We analyse TCP throughput in terms of transmission rounds, each round starting with the sender transmitting the first segment in a window of size $cwnd(t)$ at time $t$. Each round ends when the sender receives an ACK for one of the segments in a window. In this model, the duration of a round is the round trip time (RTT), and is independent of the window size. Notice that at any time $t$ holds that the number of 'pending' packets in transit at time $t$ is smaller (or equal) to congestion window size at time $t$ (unless the server is in 'fast recovery' phase at time $t$).

## 3 Motivating Anti-Replay Window

IPsec standards, [24], require anti-replay mechanism to identify and discard replayed packets in order to prevent DoS attacks, e.g., [39], claim that the reason for anti-replay mechanism is to save CPU cycles which will be wasted on replayed packets, as well as to prevent incorrect billing information. Yet to obtain access to a service or resource, the attacker will have to obtain the secret keys used to encrypt (and possibly authenticate) the communication, and it will not gain much by merely replaying already sent messages. In addition, typically, replay-sensitive applications check for freshness of messages and discard (or ignore) replayed messages. We present an additional motivation for IPsec anti-replay window; more specifically, we show that without IPsec anti-replay window, (amplification) degradation of service attacks on congestion control of TCP can be launched with significant performance damages. In what follows we describe the attacks that could be launched if no anti-replay mechanism were used. These attacks require merely a duplicating stealth MITM attacker (see Figure 2(a)).

### 3.1 ACKs Duplication: Stealth-MITM DoS Attack on Channel without Anti-Replay Mechanism

A client behind $GW1$ requests to download a file from server behind $GW2$, as in Figure 1(a). The attacker is presented in Figure 4. The main idea of the attack is to duplicate a legitimate ACK sent by the client in response to some segment, and retransmit three duplicate copies of that ACK. TCP considers the receipt of three duplicate ACKs as an indication of lost segment, which in turn can be a sign of congestion. As a result, TCP at the sender halves its congestion-control window. Furthermore, if prior to the attack the TCP connection were in 'slow start' phase (where the congestion window grows exponentially), TCP also moves to the linearly-growing 'congestion avoidance' (CA) mode, thus prematurely aborting the slow-start phase. As a result, the connection uses a small congestion window, which results in severe bandwidth underutilisation. By repeating this attack periodically, the attacker can ensure that the connection continuously uses very small, suboptimal window. In Figure 4 we present an attack on a TCP connection in CA mode. For simplicity assume that the congestion window at the beginning of first attack epoch at time $t_0$ is $cwnd(t_0) > 4 * MSS$, i.e., adversary did not inject any duplicate packets in the interval $(t_0 - T, t_0)$ (and hence can send three packets at any interval beginning from $t_0$). Assume that the server sends a window of $k+1$ segments $i, ..., i+k$, and the client upon receipt, transmits $k+1$ ACKs such that ACK on segment $i+k$ is last in the window. The attack begins when the attacker creates three duplicate copies of a last ACK (for segment $i+k$) sent by the receiver in recently transmitted window of ACK segments. At time $t_0$ (in Figure 4) the server receives three duplicate ACK copies injected by the attacker. Since we ignore transmission delays, three duplicate ACKs arrive at the same time, with no other ACK segment arriving between the most recently transmitted legitimate ACK and the receipt of three duplicates of that ACK, that were injected by the attacker. This deceives the server into believing that the three duplicate ACKs are transmitted as a result of a lost segment in the last transmitted window. The receipt of three consecutive ACKs for segment $i+k$ is taken as an indication of congestion that resulted in loss of segment $i+k$. As a result, once the server receives three consecutive duplicate ACKs, (according to step (2.d) in Figure 3, if the TCP at the sender is in CA, or step (1.e) if the TCP is in slow-start) it performs fast retransmit of the segment which it believes to have been lost (step (2.d), Figure 3), i.e., transmits the 'lost' segment for which duplicate ACKs were generated, and sets the congestion window to $cwnd(t_0^+) = \frac{cwnd(t_0^-)}{2} + 3$. The server then enters fast recovery mode (step (3.a), Figure 3) until receipt of an ACK for new data (i.e., on segment $i+k+1$), in Figure 4. Since triple duplicate ACKs were not generated due to congestion, ACK segments for all pending segments at time $t_0$ eventually arrive at the server. Once ACK acknowledging new data arrives, the congestion window is deflated (step (3.c), Figure 3), i.e., set to half of its value before the receipt of three duplicate ACKs, i.e., $cwnd = \frac{cwnd}{2}$, and the server enters congestion avoidance phase during which the sending rate grows linearly (step (2.a), Figure 3).

[2] When receiver sends an ACK for every other segment, i.e., uses delayed ACK, the congestion window grows in less than one segment per RTT; this does not significantly change our results.

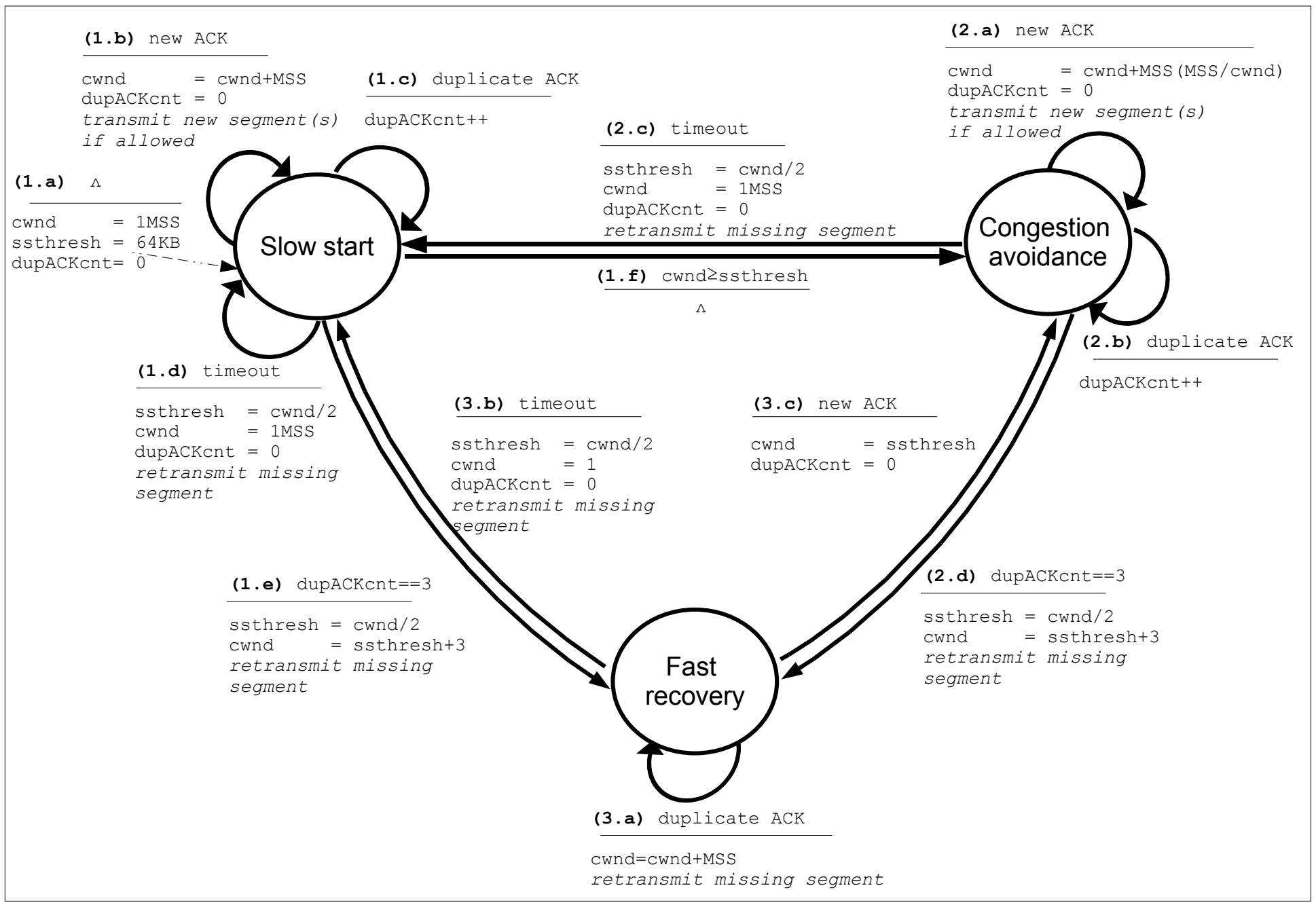

**Fig. 3.** TCP congestion control state machine of the sender (from [25]).

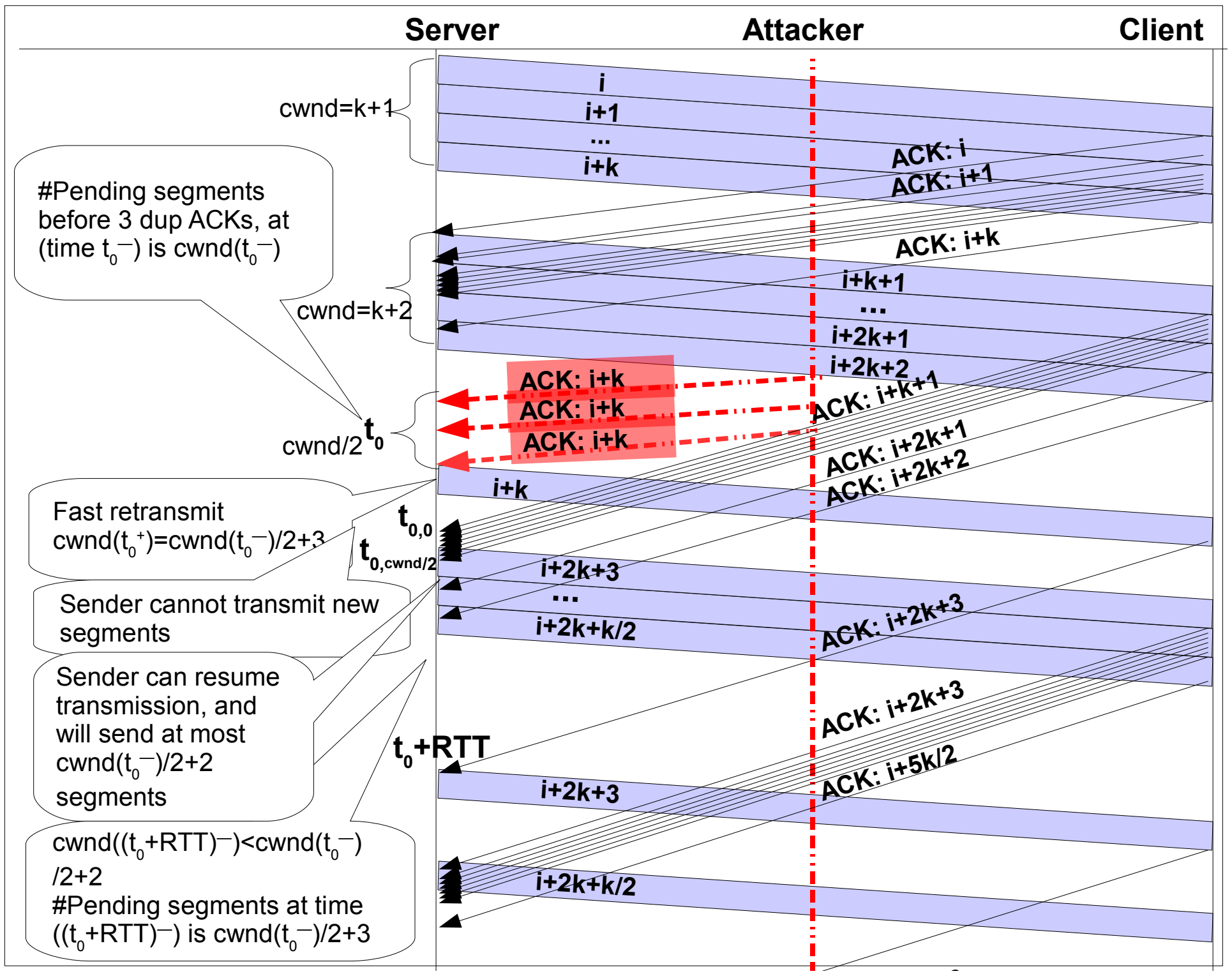

**Fig. 4.** Attack on TCP congestion control mechanism when no IPsec anti-replay window is employed. The attacker duplicates a legitimate ACK segment (sent by the client to the server in response to receipt of a data segment) and sends three duplicate copies of that ACK to the server. The server takes the three duplicate ACKs as an indication of network congestion and reduces its sending rate.

**ACKs Duplication Attack Analysis** We consider a $(\rho, 3)$-limited duplicating MITM adversary (see Figure 2(a)), with constant delays; notice we require $\sigma = 3$ since the attack requires the adversary to duplicate three ACK segments. The attacker can repeat the attack once every $T = 3/\rho$ seconds; we refer to $T$ as the length of each *attack epoch*, i.e., time elapsed between two consecutive duplications of three ACK segments by the attacker[3]. We analyse the operation of TCP in the $T$ seconds from one duplication to the next.

In Claim 1, we present a computation of an upper bound on the average steady-state congestion window $cwnd^{ATK}_{MAX} \leq \frac{3T}{2RTT}$. Then in Claim 2 we derive the throughput of the connection when under attack: $\frac{3T}{2RTT^2}$. Let $T$ be the attack epoch frequency, and $T = \frac{3}{\rho}$, then for $T = RTT$ ($T$ defined in Section 2.2), the average congestion window is $3MSS$ (maximal segment size), with throughput of at most $\frac{3}{2RTT}$ for the sender. This throughput is much lower than the throughput of connection (when not under attack), which is the average TCP congestion window (that can grow up to $65KB$) divided by the round trip time. Then in Claim 3 we compute the number $i$ of attack epochs required to reach the maximal steady state congestion window to be $i < log_2(cwnd(t_0^-) - \frac{2T}{RTT} - 2) - 1$.

[3] Injecting more than 3 duplicate segments will not have the maximal amplification effect.

**Claim 1** *Steady state congestion window $cwnd^{ATK}_{MAX}$ of the TCP sender when under packet reordering attack, with network and communication model as in Section 2.3, and attacker model in Section 2.2, Figure 2(a), is $cwnd^{ATK}_{MAX} \leq \frac{3}{2RTT}$*

*Proof.* Let $t_{i,j}$ be the $j^{th}$ ACK segment received by server (in Figure 4) at time $t_i$ after attack epoch $i$ (described in Figure 4), i.e., after the receipt of three duplicate ACKs at time $t_i$. At time $t_{0,1}$, ($t_{0,1} > t_0$) first legitimate ACK segment sent by the receiver after first attack epoch, arrives. Since we assume constant delays, the first ACK arriving at time $t_{0,1}$ is essentially a legitimate ACK generated by the receiver with a higher sequence number than the ACK duplicated by the attacker. As a result the server will exit fast recovery, will set the congestion window to $cwnd(t_{0,1}) = \frac{cwnd(t_0^-)}{2}$ thus deflating the congestion window, and will enter a CA phase during which the congestion window increases linearly, approximately by one segment during each RTT; analysis of linear congestion window growth is presented in Figure 5(a), and holds: $cwnd(t + RTT) < cwnd(t) + 1$. At time $t_{0,1}$ (when first legitimate ACK arrives) the server cannot transmit new segments into the network since $cwnd(t_{0,1}) < cwnd(t_0^-)$, where $cwnd(t_0^-)$ is the number of pending segments and $cwnd(t_{0,1}) = \frac{cwnd(t_0^-)}{2}$, i.e., the congestion window does not allow transmission of new segments. Let $n$ be the number of pending segments at time $t_0$, then $n$ ACKs should arrive at the sender at times $t_{0,1}, ..., t_n$ at constant intervals (since we assume constant delays). After arrival of $\frac{n}{2} = \frac{cwnd(t_0^-)}{2}$ ACK segments, $t_{0,1}, ..., t_{0,\frac{n}{2}}$, the server can resume transmitting segments with each new ACK arrival, and upon receipt of $cwnd(t_0^-)$ ACK segments, a total of $\frac{cwnd(t_0^-)}{2}$ segments will be transmitted. After RTT seconds, at time $t_0 + RTT$ the ACK on the retransmitted segment at time $cwnd(t_0^+)$ arrives, and the congestion window size is

$$cwnd((t_0 + RTT)^+) = \frac{cwnd(t_0^-)}{2} + cwnd(t_0^-)\frac{1}{\frac{cwnd(t_0^-)}{2}} = \frac{cwnd(t_0^-)}{2} + 2$$

Next attack epoch is initiated $T = 3/\rho$ seconds later, i.e., it takes $T = 3/\rho$ seconds until attacker can launch the attack again, by sending three subsequent duplicate ACK segments. Since the connection at this time is in congestion avoidance (CA) phase, i.e., congestion window grows by one segment in every round-trip time (RTT), the congestion window at second attack epoch, at time $t_1 = t_0 + \frac{T}{RTT}RTT$ will have reached $cwnd(t_1) = \frac{cwnd(t_0^-)}{2} + \frac{T}{RTT}$, where $\frac{T}{RTT}$ is the number of RTTs between each attack epoch. Let $t_i = t_0 + i * 3/\rho = t_0 + i\frac{T}{RTT}RTT$ for $i \in \mathbb{N}$, be the time at $i^{th}$ attack epoch. We derive the congestion window size $cwnd$ at time $t_i = t_0 + i\frac{T}{RTT}RTT$ as a function of the frequency of the attack and of the congestion window $cwnd(t_0)$ (the window prior to first attack epoch on the connection).

$$cwnd(t_i) = \frac{cwnd(t_{i-1})}{2} + \frac{T}{RTT} = \frac{cwnd(t_0^-)}{2^i} + \frac{(2^i - 1)}{2^{i-1}}\frac{T}{RTT} = \frac{2T}{RTT} + \frac{1}{2^i}\Big(cwnd(t_0^-) - \frac{2T}{RTT}\Big) \quad (1)$$

Since $\frac{1}{2^i}$ is negligible, the whole expression is approximates $\frac{2T}{RTT}$, i.e., $cwnd(t_i) \leq \frac{2T}{RTT}$. For $T = 3/\rho$, $cwnd(t_i) \leq \frac{6}{\rho \cdot RTT}$. In addition, since for every $t_i$, $cwnd^{ATK}_{MAX} \leq cwnd(t_i)$, the bound on the steady state congestion window size when under attack is given by: $cwnd^{ATK}_{MAX} \leq \frac{6}{\rho \cdot RTT}$ The average congestion window size at steady state when connection is under attack is presented in Figure 5(b) (note that the growth of window size is descerete, i.e., upon each packet arrival). This window size results in data transfer rate of at most $\frac{cwnd^{ATK}_{MAX}}{RTT} = \frac{6}{\rho}$, which can be very small - in fact, negligible compared to the expected throughput without attack. □

**Claim 2** *The throughput of average steady state congestion window is at most $\frac{3T}{2RTT^2}$.*

*Proof.* To compute throughput we take the long term data exchange for the average window size derived in Claim 1, and divide it by the RTT:

$$\lim_{i \to \infty} \frac{\frac{cwnd(t_0^-)}{2^i} + \frac{(2^i - 1)T}{2^{i-1}RTT}}{RTT} = \frac{2T}{RTT^2} \leq \frac{6}{\rho \cdot RTT^2}$$

**Claim 3** *Let $i$ be the number of attack epochs required to reach the steady state at time $t_i$ with window size of $cwnd(t_i^-)$, then $i < log_2(cwnd(t_0^-) - \frac{2T}{RTT} - 2) - 1$.*

*Proof.* Assume at time $t_i$, steady state is reached, and according to Claim 4, $cwnd(t_i) - 1 < cwnd(t_{i+1})$; then $cwnd(t_i) < \frac{2T}{RTT} + 1$. In claim 5 we derived the expression for congestion window to be $cwnd(t_i^-) = 2\Big(\frac{T}{RTT}+1\Big) + \frac{1}{2^{i+1}}\Big(cwnd(t_0^-) - 2(\frac{T}{RTT}+1)\Big)$; by substituting $cwnd(t_i)$ with $\frac{2T}{RTT}+1$ we obtain $cwnd(t_i^-) = 2\Big(\frac{T}{RTT}+1\Big) + \frac{1}{2^{i+1}}\Big(cwnd(t_0^-) - 2(\frac{T}{RTT}+1)\Big) < \frac{2T}{RTT}+1$; by solving the equation we obtain that upper bound on the number of attack epochs to reach steady state is $i < log_2(cwnd(t_0^-) - \frac{2T}{RTT} - 2) - 1$. □

**Claim 4** *The steady state congestion window $cwnd(t_i)$ is reached during attack epoch $t_i$, when $cwnd(t_i) < cwnd(t_{i+1}) + 1$, in attack epoch $i$ at time $t_i$.*

*Proof.* We consider the following two cases related to congestion window growth between each subsequent attack epoch:

$\frac{cwnd(t_i)}{2} > \frac{T}{RTT}$ The window size decreases between each subsequent attack epoch; this is due to the fact that the duration between attack epochs does not suffice for congestion window to restore its size.

$\frac{cwnd(t_i)}{2} \leq \frac{T}{RTT}$ The congestion window size is restored between each attack epoch.

Conside the case where $cwnd(t_i) > \frac{2T}{RTT}$; the congestion window size decreases between each two subsequent attack epochs $t_i$ and $t_{i+1}$: at each attack epoch holds: $cwnd(t_{i+1}) + 1 \leq cwnd(t_i)$. Once $cwnd(t_i) \leq \frac{2T}{RTT}$, the congestion window between attack epochs $t_i$ and $t_{i+1}$ is: $cwnd(t_i) < cwnd(t_{i+1}) + 1$, and the steady state window size is reached; then $cwnd(t_{i+1}) \leq cwnd(t_i) < cwnd(t_{i-1})$, where $cwnd(t_{i-1})$ is congestion window right before steady state, and $cwnd(t_i) < cwnd(t_{i+1}) + 1$. □

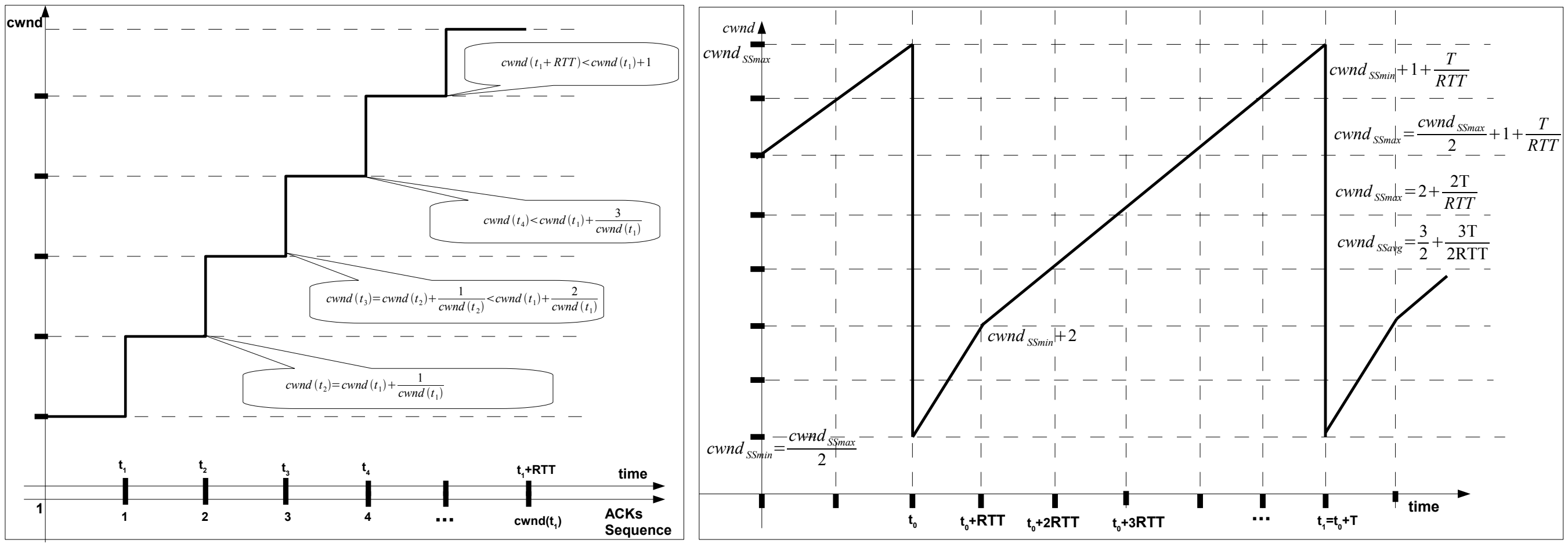

(a) TCP congestion window growth upon receipt of each new ACK, when in congestion avoidance phase.

(b) Average congestion window $cwnd$ at steady state when connection is under packet-reordering attack. When $T = RTT$ average $cwnd = 3$ and throughput is at most $\frac{3}{RTT}$.

**Fig. 5.** Attack on TCP congestion control mechanism when no IPsec anti-replay window is employed. The attacker duplicates a legitimate ACK segment (sent by the client to the server in response to receipt of a data segment) and sends three duplicate copies of that ACK to the server. The server takes the three duplicate ACKs as an indication of network congestion and reduces its sending rate. The attack is launched in slow start in (4).

### 3.2 Data Packets Duplication: Stealth MITM DoS Attack on Channel without Anti-Replay Mechanism

The attack on data packets duplication, albeit similar in nature to attack on ACKs, serves several purposes: it may be the case that the attacker can only duplicate and inject packets in one direction, in which the packets flow and not in the direction in which the ACKs flow, i.e., can inject segments sent from server to client but cannot inject segments from client to server, e.g., if the attacker is located in one network and can only duplicate and inject packets in that same network, or if there is an ingress filtering mechanism. Another purpose is related to assumptions on the power of the attacker: the duplicating packets attack can be carried out by a much weaker (slower), attacker, that injects segments that are much older than the legitimate segments. This attack will not succeed with duplicating ACK segments, since the sender will ignore outdated duplicate ACKs, i.e., if a new ACK has already arrived the sender ignores the old ACK that arrives. The attack is symmetric to the duplication of ACKs presented above, and due to space restrictions we only briefly sketch an outline of the idea. In order to achieve performance reduction, the attacker tricks the receiver into generating duplicate ACKs by duplicating packets. Attacker injects a three duplicate copies of a previously sent packet, which triggers three duplicate ACKs at the receiver. As defined in [18], the TCP receiver acknowledges every packet received. After three consecutive duplicate ACKs the sender performs fast retransmit and slows down its sending rate.

## 4 Determining Size of IPsec Anti-Replay Window

The attacks presented in previous section motivate the necessity for anti-replay mechanism, that prevents duplicates since replayed packets can be exploited by malicious attacker to degrade performance by launching an amplification DoS attack. The standard [23], recommends using anti-replay mechanism as a protection against denial of service (DoS) attacks by a MITM adversary, however it does not specify how to calculate the proper window size, and there is no analysis of anti-replay mechanism in malicious setting. Existing works [22, 15, 39] recommend using anti-replay mechanism against a MITM adversary to prevent replay of messages, and propose efficient implementations of anti-replay window. Yet 'resource efficient' implementations may expose the IPsec implementation to DoS attack as we will show in this section. In this section we show that if the anti-replay window is not correctly adjusted, a more severe performance damage could be induced to the TCP flows, than not using the anti-replay window at all. We then show how to calculate the correct window size in Section 4.3, given the relevant network parameters.

### 4.1 Single Flow Security Association (SA) vs. Multiple Flow Security Association (SA)

The standard [23] specifies establishing a security association (SA) in each direction between the communicating parties, but it does not specify whether to establish a single SA in each direction or if to establisha distinct SA for each traffic flow. A single SA for all flows is more efficient since the gateway is required to keep less state. In addition, single SA prevents distinguishing between different communication flows. On the other hand it increases the risk for a chosen plaintext attack (CPA), [4], since the attacker can 'use' other flows to attack a target flow (thus for cryptographic purposes it is recommended to use a distinct SA per each flow). Furthermore, when a single SA is used for all flows our attack on small anti-replay window is more devastating, and the attack can be launched more frequently. An SA per each flow allows distinguishing between different flows, albeit when one flow is attacked, others are not effected. On the flip side, a distinct SA for each connection imposes a significant overhead, and requires larger memory resources. However, even when a single SA pair per connection is established the reduction of performance attack is still possible, albeit not as effective as compared to a single SA pair.

### 4.2 Packets Speedup: Stealth-MITM DoS Attack on Small IPsec Anti-Replay Window

Packet reordering occurs when packets are received in a different order from the one in which they were sent. Packet reordering can significantly degrade TCP performance. Upon duplicate ACKs the sender triggers *fast* *retransmit* and *fast recovery*. As a result the congestion window remains small relatively to the available bandwidth. As specified in [3], out of order data segments should be acknowledged immediately in order to accelerate loss recovery. To trigger the fast retransmit algorithm the receiver should send an immediate duplicate ACK when it receives a data segment above a gap in the sequence space. If messages are reordered on transit, e.g., due to benign network reordering, or by malicious attacker, there will be a gap in sequence numbers of the arrived packets, which may result in anti-replay mechanism discarding valid messages if the anti-replay window is incorrectly adjusted, i.e., too small. Note that in our attack we assume that the flow control does not restrict the congestion window growth.

If a single security association (SA) pair is established for all communication, then the attacker can mount the attack more frequently. Since we give upper bounds, we assume worst case, and thus throughout the attack we assume a single TCP flow which the attacker attempts to attack, with no other communication (this is equivalent to using a distinct SA per connection). The attack can be launched when IPsec anti-replay window is smaller than the TCP congestion window. The IPsec anti-replay window size is part of the design and is known to the attacker, and due to the fact that the attacker is eavesdropping on the communication and knows the TCP window size.

An adversary can cause an IPsec implementation to discard valid packets by injecting replayed packets with higher sequence number into the message stream thus advancing the anti-replay window, and as a result legitimate packets with low sequence numbers, i.e., to the left of the anti-replay window, will be discarded by the IPsec. Discarded packets result in three duplicate ACKs at the sender, which then reduces the TCP congestion window. The throughput of TCP connections under attack is significantly degraded. The damage is a function of the frequency at which the adversary can launch the attack. The attacker model we consider in this section is presented in Figure 2(b).

**Attack** The attack presented in Figure 6(a). Assume that IPsec anti-replay window consists of $W = n$ packets and TCP window is comprised of $cwnd = k + 1$ segments, such that $n < k + 1$. The sender receives $k$ ACKs and transmits a window of $k + 1$ segments, with $i^{th}$ segment being the first segment in the TCP window and $i + k$ being the last segment, with highest sequence number.

The attacker duplicates and injects a segment[4] (speeding it up) with sequence number $i+n$ (segment $n+1$ in the TCP window), which is the first segment to the right of the anti-replay window, such that it arrives before the first segment $i$ in the window. Thus upon receipt of the segment, IPsec implementation in gateway $GW1$ advances the anti-replay window with segment $i + n$ being the right edge of the window, and passes this segment to the receiver. When the TCP at the client receives a segment with sequence number $i + n$ it generates a duplicate ACK with sequence number $i$, i.e., sequence number of the expected segment, indicating a gap in sequence numbers of the received segment. The rest $k$ segments arrive intact, and are passed to the client. When the original segment with sequence number $k + 1$ arrives, anti-replay detects a replay and thus discards it. For each subsequently received segment, with sequence numbers in the range between $i + 1$ and $i + n - 1$, the TCP at the client generates a duplicate ACK, indicating that it is missing a segment with sequence number $i$. A total of $k$ duplicate ACKs are returned to the sender. The segments loss is taken as an indication of congestion and an indication that the TCP window has grown larger than the network can handle, hence TCP takes corrective steps by decreasing the window. Denote by $t_0$ the time at which the sender receives three duplicate ACKs. According to the specification in RFC 2581 [2], upon receipt of 3 duplicate ACKs, the fast retransmit algorithm at the sender retransmits the lost segment, i.e., segment with sequence number $i$, the sender halves the congestion window and adds the three duplicate ACKs, resulting in $cwnd(t_0) = \frac{cwnd(t_0^-)}{2} + 3$. Then the sender enters a fast-recovery phase until a non-duplicate ACK arrives. At this stage the congestion window does not allow transmission of new segments into the network, since the number of pending segments is larger than the congestion window size, i.e., $cwnd(t_0^-) + 1 > \frac{cwnd(t_0^-)}{2} + 3$. For each subsequent duplicate ACK the sender increments the congestion window by 1 MSS. The addition of duplicate ACKs to the congestion window artificially inflates the window by the number of segments that have left the network and arrived at the receiver. After receipt of $\frac{cwnd(t_0^-)}{2} + 1$ duplicate ACKs (since $t_0$) the sender can resume transmission of new segments; and $\frac{1}{2}cwnd(t_0^-) - 2$ remaining duplicate ACKs will arrive. For each duplicate ACK the sender transmits a new segment into the network. In this transmission round a total of $\frac{1}{2}cwnd(t_0^-)$ segments will be sent.

[4] In this attack we assumed that the attacker speeds-up a single segment. If attacker speeds up $l$ segments, then the 'recovery' is much slower, since each of the $l$ segments will have to be retransmitted, and thus recovery will require additional $l$ transmission rounds.

At time $t_0 + RTT$ the sender should receive an ACK for the retransmitted segment. Once the server receives an ACK for new data[5] the congestion window is deflated, i.e., set to $\frac{cwnd(t_0^-)}{2}$, and the sender enters a congestion avoidance (CA) phase, during which the congestion window is incremented linearly, i.e., roughly by one segment every RTT.

Note that for this attack to succeed, we assume that the IPsec anti-replay window is smaller than the TCP congestion window, which is a reasonable assumption since provided no network congestion, the TCP window is limited only by threshold, which is typically set to 65KB, according to RFC 2988 [31].

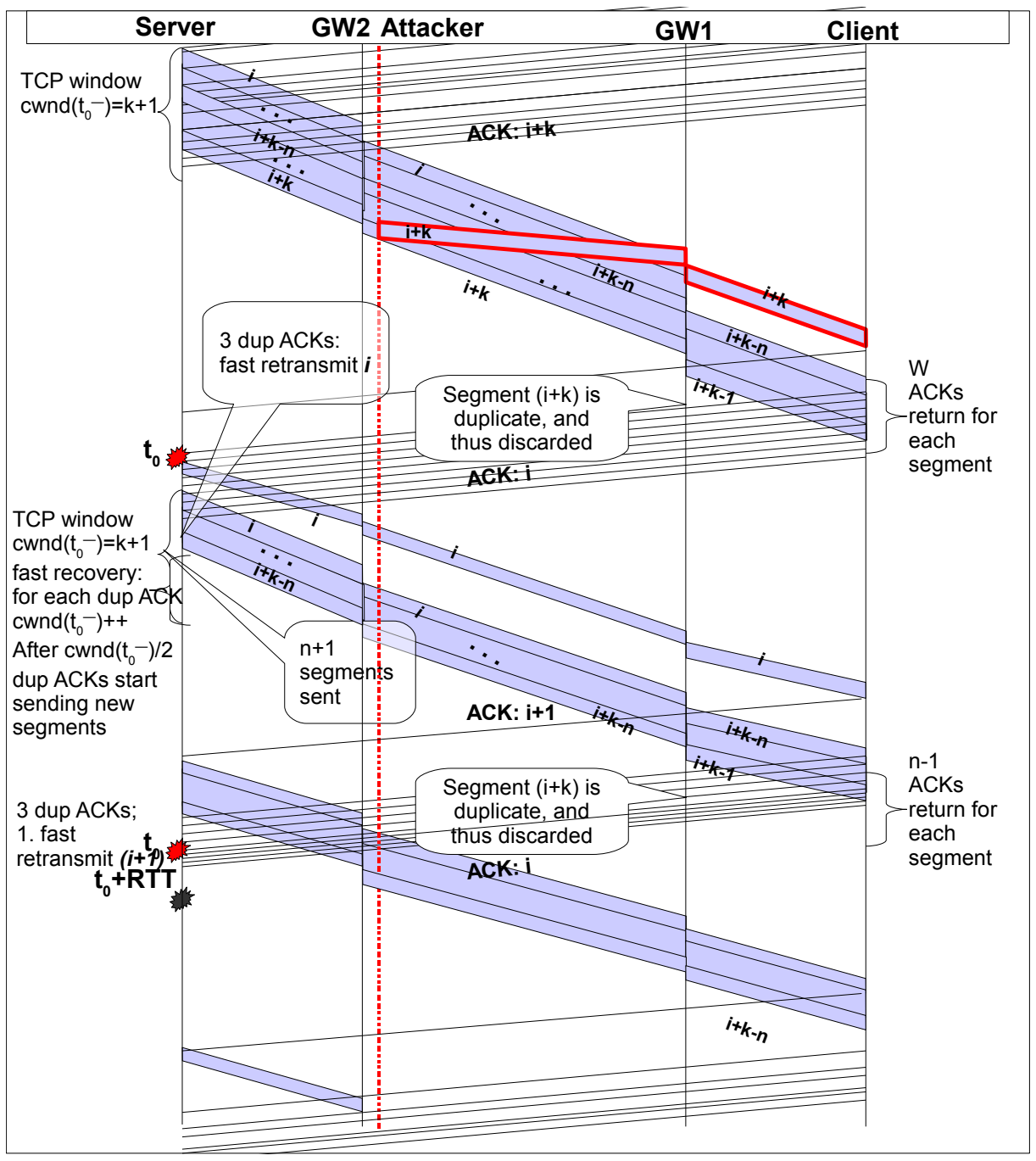

(a) Packet speedup attack on TCP exploiting an insufficient size of IPsec anti-replay window; single attack epoch.

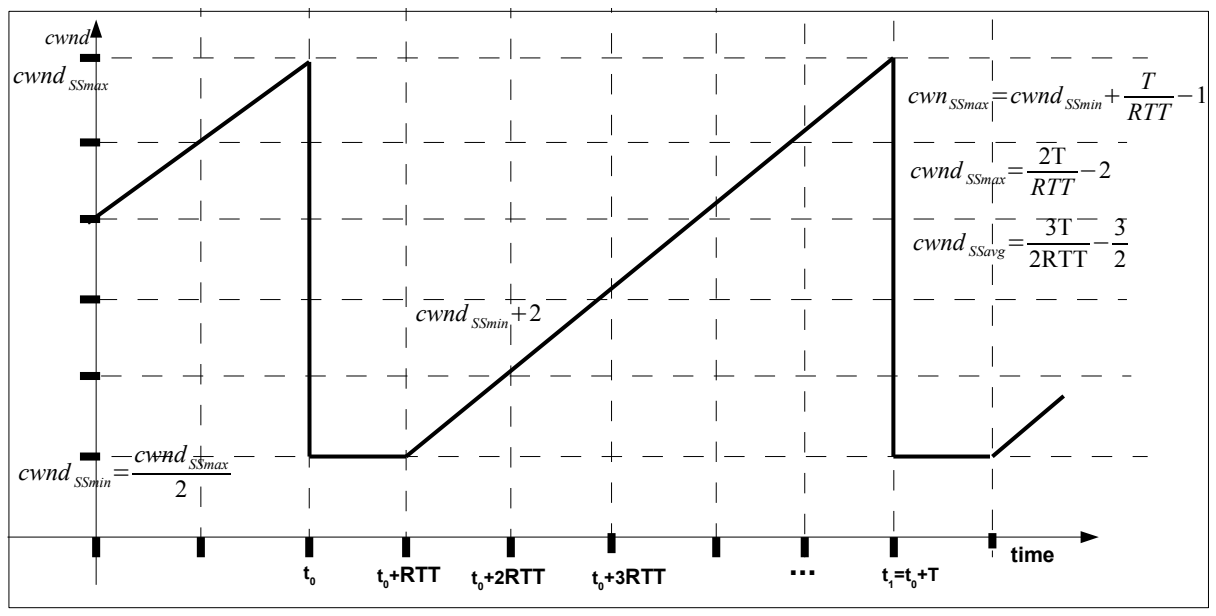

(b) TCP congestion window *cwnd* analysis, when TCP connection is under stealth DoS MITM attack. .

**Fig. 6.** In Figure 6(a), we present a speedup attack, where an attacker speeds up one segment, thus advancing the IPsec anti-replay mechanism, with this duplicate segment being the right edge. As a result, a segment with lower sequence number, to the left of the anti-replay window is discarded, resulting in a window of duplicate ACKs sent by the receiver. In Figure 6(b) we analyse the congestion window of TCP when the connection is under speedup duplication attack.

[5] We stress, that for a larger ratio of TCP congestion window and IPsec anti-replay window, a more devastating attack is possible. More specifically, the attacker will again speed-up the retransmitted segment $i$, which will again be discarded, and thus the sender will continue receiving duplicate ACKs till it encounters a timeout event. After a timeout the sender again retransmits the 'lost segment $i$' and enters a slow start. However, then it receives duplicate ACKs for $i^{th}$ segment from previous transmission round, and enters congestion avoidance. If the *cwnd* vs. $W$ (IPsec anti-replay) is sufficiently large, and enough duplicate ACKs return, the connection will eventually be reset.

**Packets Speedup Attack Analysis** The throughput of the connection is kept low, since the adversary can resume the attack (if the attack frequency parameter allows it) every time the congestion window is larger than the anti-replay window. The ratio between the TCP window *cwnd* and the IPsec anti-replay window $W$ before the first attack epoch, as well as the frequency at which the attacks can be launched, dictates the performance degradation inflicted by the attack, and the impact can range between degradation of service and acomplete denial of service. If TCP congestion window is larger than IPSec anti-replay by 1 segment, then attack achieves a result similar to reduction of quality (RoQ) attacks, in [16, 17]. In this case, it will take $\frac{k+3}{2}$ RTTs to restore the congestion window from $\frac{k-1}{2}$ back to its original value, before the first attack, i.e., $k+1$, since in every RTT the congestion window grows by one segment. Alternately, if $cwnd \geq 2*W+4$ attacker can disrupt the connection by causing the retransmission timeout (RTO) to expire, thus performance degradation induced by the attack is similar in its result to the low rate attacks presented in [26]. The minimal *cwnd* and $W$ ratio which would result in timeout of the retransmitted (due to receipt of three duplicate ACKs) segment $i$, should be computed as follows: let *cwnd* be the amount of transmitted segments in the window prior to first attack epoch. Denote by $P$ the number of segments in transit (for which the sender has not yet received acknowledgments), and denote by $W$ the anti-replay window size. In order for the attack to result in a timeout the following inequality has to satisfy: $\lfloor \frac{cwnd}{2} \rfloor + (cwnd - 1) - P > W$. The number of pending segments $P$ is equal to the number of transmitted segments, i.e., $P = cwnd$, therefore in order for the attack to result in a timeout, the *cwnd* and $W$ ratio has to satisfy: $\lfloor \frac{cwnd}{2} \rfloor - 1 > W$. The throughput analysis of the attacked connection is in Figure 6(b).

### 4.3 Solutions and Adjusting IPsec Anti-Replay Window

In order to prevent the denial/degradation of service attacks we presented, a larger anti-replay window should be used, and the question is how much larger. The largest possible IPsec anti-replay window is the one that can contain all the possible packets within a specific SA, i.e., window of size $2^{32}$ bits, which obviously prevents the attacks which were possible on small window, i.e., even severe reordering of segments will not result in dropped packets. However, an anti-replay window of size $2^{32}$ is inefficient and impractical. Existing works [39, 15, 22], investigate constructions of optimal anti-replay mechanism, to reduce the resources required to maintain the anti-replay window.

We next compute the upper bound on the number of packets that the anti-replay window should reflect, in order to prevent reordering attacks (which result in denial of service), when small anti-replay window is used. IPsec anti-replay window size should be computed based on the rates of the given network. When distinct SAs are used for each TCP flow, IPsec window should be as large as the upper bound on the size of the TCP window. Alternately, when a single SA pair is used for all traffic, IPsec window computation should be a function of the network parameters, i.e., the transmission rate, and speed on the channels between the virtual private networks (VPNs), as well as maximal possible delay in a network.

Let $L$ be the maximal packet size, and let $R$ be the transmission rate in bytes per second (Bps). Then the number of packets in transit is $N$, s.t. $N = \frac{R \times d_{PROP}}{L}$, where $d_{PROP}$ is the maximal delay, i.e., if a packet does not arrive within $d_{PROP}$ seconds it is assumed to have been lost. The IPsec window should be at least the size of the maximal number of packets in transit at a given time. Assume packet size of $L = 10^3$ bytes, $d_{PROP} = 1$ second and $R = 10$ MBps, then the maximal number of packets in transit is $10,000$. This prevents the attacker from discarding out of order packets, by advancing IPSec anti-replay window.

Note that this is a rather conservative computation, since typically, the attacker speed will also be a function of the delay, i.e., even the attacker does not have 0 seconds delay. However, since we compute upper bounds we assume worst case scenario. We differentiate between maximal window size $W$, i.e., the number of bits required to maintain the anti-replay window, and the number of packets $N$ that the window of size $W$ can contain. A trivial anti-replay window size $W$ for $N$ packets is $W = N$. While this lower bound on the anti-replay window will prevent the speedup attacks which advance the anti-replay window and result in discarded segments, it can be too large and inefficient for practical purposes, i.e., maintaining such a large window can be a challenge, w.r.t. processing requirements and storage resources (especially if a distinct SA pair is established for each flow). Thus the goal is to maintain a window of size $W < N$ where $N$ is the number of packets that the window maintains. Usually the anti-replay window is a sparse vector with out of order packets, and allocating large buffers may be alleviated with a more efficient data structure. Naive solutions that attempt to save resources by decreasing window size are susceptible to attacks, and may result in more damage compared to IPsec with no anti-replay mechanism. There are works that attempt to achieve a more efficient and robust anti-replay window management, e.g., [22, 23, 15]. However, there are no works that analyse the anti-replay window in an adversarial setting, where attacker can maliciously adjust its strategy. We leave it as an open question to come up with an efficient and robust anti-replay window design in an adversarial setting.

## 5 IPsec with Large Anti-Replay Window

If IPsec anti-replay window is not adjusted properly, i.e., too small, reordering of packets (e.g., by malicious attacker) can degrade performance of TCP connections. However, as we show in this section, even sufficiently large IPsec anti-replay window does not prevent throughput degradation attacks. In this section we assume that there are available resources to maintain an anti-replay window of the required size to prevent attacks that advance the anti-replay window resulting in lost packets. We consider a $(\rho, 3)$-limited stealth MITM attacker, following the model in Figure 2(c). Note, that this is a similar attacker model as the one assumed in the attack in Section 3, with an additional capability to speed-up network packets, i.e., an attacker can send its packets via a faster route than the one available to the legitimate parties. We show that the throughput degradation is identical to the one in Section 3, which proves that IPsec anti-replay mechanism does not prevent degradation-of-service attacks, even when the anti-replay mechanism is of sufficient size.

### 5.1 Packets Reordering: Stealth-MITM DoS Attack on Channel with Infinitely Large Anti-Replay Window

Consider the scenario in Figure 1(a), in which client behind gateway $GW1$ requests to download a file from server behind gateway $GW2$, and assume IPsec implementation is using sufficiently large IPsec anti-replay window. We present a detailed attack on this connection in Figure 7(a). The server transmits a window of $k+1$ segments to the client, and assume $k \geq 3$. Attacker speeds up three segments[6] with higher sequence numbers, i.e., $(i+k-2), (i+k-1), (i+k)$. These segments arrive before $i^{th}$ segment to the gateway. IPsec authenticates the segments, and verifies that they are not a replay of previously sent segments and passes them to the client. When the client receives these out of order segments, it detects a gap in sequence numbers of received packets, i.e., the sequence number is higher than the one expected. As a result, the client generates a duplicate acknowledgment for next segment it expects to receive, i.e., segment with sequence number $i$. Once the sender receives three duplicate ACKs for the same segment, it retransmits the 'lost' segment, halves the congestion window, and enters fast recovery (step (2.d) if TCP is in CA or step (1.e) if TCP is in slow start; in Figure 3). Upon arrival of an ACK acknowledging new data, i.e., ACK for segment $(i+1)$, (step (3.c) in Figure 3) the congestion window is deflated, i.e., set to half of its value before the receipt of three duplicate ACKs, and the server enters congestion avoidance phase, i.e., increases the congestion window by roughly 1 segment in every RTT (step (2.a) in Figure 3). Note: no packets are discarded by IPsec implementation, since the anti-replay window is assumed to be sufficiently large.

### 5.2 Packet Reordering Attack Analysis

In Claim 5 we show that an average congestion window at steady state when under attack is $cwnd^{ATK}_{AVG} \leq \frac{3T}{2RTT}$, with an upper bound on the maximal steady state congestion window for the sender of $cwnd^{ATK}_{MAX} \leq \frac{2T}{RTT}$. We then derive, in Claim 6, the long-term throughput of the connection when under packet reordering attack: $\frac{3T}{2RTT^2}$. Let $T$ be the attack epoch frequency, and $T = \frac{3}{\rho}$, then for $T = RTT$ ($T$ defined in Section 2.2), the resulting average congestion window is 3 MSS, with throughput of at most $\frac{3}{2RTT}$. Note that this throughput is much lower than the average throughput of TCP connection that is not under attack, i.e., throughput is the average TCP congestion window (which can be up to $65KB$ divided by the round trip time (RTT)). Then in Claim 7 we compute the number $i$ of attack epochs required to reach the maximal steady state congestion window to be $i < log_2(cwnd(t_0^-) - \frac{2T}{RTT} - 2) - 1$.

[6] Similarly to attack in Section 3, optimal amplification impact is achieved when attacker speeds up exactly three segments. If less than three packets arrive out of order, the sender will not reduce the congestion window but instead will wait for a timeout. However, since the segments transmitted by the sender eventually arrive, and ACKs for them are returned, no timeout will occur.

**Claim 5** *Steady state congestion window $cwnd^{ATK}_{MAX}$ of the TCP sender when under packet reordering attack, with network and communication model as in Section 2.3, and attacker model in Section 2.2, Figure 2(c), is* $cwnd^{ATK}_{MAX} \leq \frac{3}{2RTT} + 2$

*Proof.* Assume that at time $t_0$ the sender is in slow start or in congestion avoidance state (in Figure 3) and it receives three duplicate ACKs. Let $t_0^-$ be the time right before arrival of three duplicate ACKs, and $t_0^+$ be the time right after the arrival of three duplicate ACKs, and denote by $cwnd(t_0^-)$ the congestion window size at time $t_0^-$ (this is also the number of ACKs that will arrive following time $t_0^-$). For each duplicate ACK the *dupACKcnt* variable is incremented, and once $dupACKcnt = 3$, the TCP at the sender transitions to fast recovery (step (1.e) in Figure 3): sets the *ssthresh* to $\frac{1}{2}cwnd(t_0^-)$, and sets the congestion window to $cwnd(t_0^+) = ssthresh + 3$, fast retransmits the 'missing segment', and switches to fast recovery. When an ACK (for new data) at time $t_{0,1}$ arrives (phase (3.c) in Figure 3), the sender sets the congestion window to $cwnd((t_{0,1})^+) = ssthresh$, sets $dupACKcnt = 0$, and transforms to congestion avoidance. The number of pending (transmitted but not yet ACKed) segments at time $(t_{0,1})^+$ is: $cwnd(t_0^-) + 1$, which is greater than the congestion window size: $\frac{1}{2}cwnd(t_0^-)$, therefore, according to Figure 3, the sender cannot transmit new segments into the network (the sender can transmit segments when the amount of transmitted but yet to be ACKed (a.k.a. pending) segments is less than the size of the congestion window). Below is an expression to calculate the number $N_A$ of ACKs required for the sender to resume transmission of new data segments: $\frac{1}{2}cwnd(t_0^-) + N_A = cwnd(t_0^-) + 1 \Longrightarrow N_A = \frac{1}{2}cwnd(t_0^-) + 1$. Namely, the sender can resume transmission after receipt of $\frac{1}{2}cwnd(t_0^-)+1$ ACKs, i.e., ACKs arriving at: $t_{0,1}, ..., t_{0,k}$, for $k = \frac{1}{2}cwnd(t_0^-)+1$. The number of remaining ACKs to arrive is $\frac{1}{2}cwnd(t_0^-) - 5$, i.e., ACKs that will arrive at time: $t_{0,k+1}, ..., t_{0,2k-1}$. Since in congestion avoidance the congestion window grows linearly, roughly by $MSS\frac{MSS}{cwnd}$ for each ACK (according to step (2.a) in Figure 3), the amount of segments that the sender will transmit in $[t_0^+, (t_0 + RTT)^-]$ (and the congestion window size at time $(t_0 + RTT)^-$) is given by

$$cwnd((t_0 + RTT)^-) < \frac{1}{2}cwnd(t_0^-) + (cwnd(t_0^-) - 4)\frac{1}{\frac{1}{2}cwnd(t_0^-)} = \frac{1}{2}cwnd(t_0^-) + 2 - \frac{8}{cwnd(t_0^-)}$$

Since $\frac{8}{cwnd(t_0^-)} > 0$ the number of transmitted segments in interval $[t_0^+, (t_0+RTT)^-]$ is at most $\frac{1}{2}cwnd(t_0^-)+2$; and $cwnd((t_0+RTT)^-) < \frac{1}{2}cwnd(t_0^-)+2$. At time $t_0+RTT$ an ACK for the segment (retransmitted at time $t_0^+$) should arrive (since we assume no loss and constant delays), followed by at most $\frac{1}{2}cwnd(t_0^-) + 2$ ACKs in response to earlier transmitted segments, i.e., the segments transmitted in interval $[t_{0,k+1}, (t_0 + RTT)^-]$. The congestion window size by the end of $(t_0 + RTT)^{th}$ transmission round (at time $(t_0 + 2RTT)^-$), is

$$cwnd((t_0+2RTT)^-) < cwnd((t_0+RTT)^-)+\frac{1}{2}cwnd(t_0^-)\Big(\frac{1}{\frac{1}{2}cwnd(t_0^-)}\Big) = \frac{1}{2}cwnd(t_0^-)+2+1 = \frac{1}{2}cwnd(t_0^-)+3$$

Namely, the congestion window increases by at most one MSS in each transmission round (RTT). By next attack epoch, at time $t_1$, the congestion window $cwnd(t_1^-)$ will have grown by at most $\frac{T}{RTT}$, and holds $cwnd(t_1^-) < \frac{1}{2}cwnd(t_0^-) + \frac{T}{RTT} + 1$ More generally, by $i^{th}$ attack epoch the congestion window $cwnd(t_i^-)$ will be:

$$cwnd(t_i^-) < \frac{cwnd(t_{i-1}^-)}{2}+\frac{T}{RTT}+1 = \frac{cwnd(t_0^-)}{2^{i+1}}+\sum_{j=0}^{i}\Big(\frac{T}{(j+1)RTT}+\frac{1}{2^j}\Big) = \frac{cwnd(t_0^-)}{2^{i+1}}+\frac{2^{i+1}-1}{2^i}\Big(\frac{T}{RTT}+1\Big)$$

$$cwnd(t_i^-) < 2\Big(\frac{T}{RTT}+1\Big)+\frac{1}{2^{i+1}}\Big(cwnd(t_0^-)-2(\frac{T}{RTT}+1)\Big)$$

The whole expression approximates $2\frac{T}{RTT}$, thus the bound on congestion window size at time $t_i^-$ is $cwnd(t_i^-) \leq \frac{2T}{RTT}$. In addition, since for every $t_i$ holds: $cwin_{ATK}^{MAX} \leq cwnd(t_i)$; the bound on the steady state congestion window size when under attack is $cwin_{ATK}^{MAX} \leq \frac{2T}{RTT}$, and average congestion window size at steady state of $\frac{3T}{RTT}$. □

**Claim 6** *The throughput of average steady state congestion window is at most* $\frac{3T}{2RTT^2}$.

*Proof.* To compute throughput we take the long term data exchange for the average window size derived in Claim 5, and divide it by the RTT:

$$\lim_{i\to\infty}\frac{2\Big(\frac{T}{RTT}+1\Big)+\frac{1}{2^{i+1}}\Big(cwnd(t_0^-)-2(\frac{T}{RTT}+1)\Big)}{RTT} \leq \frac{3T}{2RTT^2}$$

We ignore the part of the connection when the window keeps decreasing since it is negligible w.r.t. the overall connection throughput. □

**Claim 7** *Let $i$ be the number of attack epochs required to reach the steady state at time $t_i$ with window size of $cwnd(t_i^-)$, then $i < log_2(cwnd(t_0^-) - \frac{2T}{RTT} - 2) - 1$.*

The proof is the same as in Claim 3, Section 3.1.

### 5.3 Solutions: Reordering Tolerant Tunneling Protocol (RTTP)

Two independent directions can be pursued in order to address the attack in Section 5.1. One is to adjust TCP to diverse network conditions, i.e., to immune TCP from packet reordering. A wide range of TCP modifications has been proposed to improve robustness to reordering, e.g., [7, 28, 37, 38, 9, 5, 36, 10, 8]; see a survey in [27] and an analysis in [7]. The main idea of all those solutions is to detect and ignore false duplicate ACKs. Sender halves the congestion window upon duplicate ACK, but then restores it back when receiver signals receipt of 'supposedly lost' segment, thus resulting in an insignificant slowdown. Yet none of the proposed solutions is widely adopted. Changing TCP requires a change in every end host, and may take considerable time to adopt.

We propose solution in the IPsec gateway. Our solution could both address benign network congestion, as well as malicious reordering. Solution in firewall does not require changing each host separately, but only to apply the modification to the firewall, and as a result to protect subnet of hosts. Many private networks connected to the Internet are protected by firewalls. Firewall protection is based on the idea that all packets destined to hosts behind a firewall have to be examined by the firewall. The drawback of this approach is the additional processing delay on every packet, and having the firewall maintain state. On the other hand, many firewalls examine TCP connections for security reasons, e.g., solution to SYN attack, thus firewalls keep state. Therefore, we believe that our addition is minimal. Our solution is comprised of two phases: first detection, then prevention of an attack.

The main idea of our proposition is to delay the congestion response, i.e., ACK in the firewall and not deliver to the host. In our solution the sending gateway ($GW2$ in Figure 1(a)) will timestamp every outgoing packet, and the timestamp will be authenticated by IPsec. Receiving gateway ($GW2$ in Figure 1(a)) will detect suspicious packets (arriving out-of-order and faster than typical delay) and then will apply its decision. Upon receipt of a suspicious packet two possible actions can be applied. The trivial solution is to buffer suspected packets, and deliver after usual delay or when the gap is filled (i.e., the legitimate packets arrive). However note that this solution requires high memory consumption. We propose the 'aggressive solution', according to which the receiving gateway will delay duplicate ACKs of suspicious packets. Then if legitimate (believed to be lost) packet arrives, the delayed duplicate ACK will be discarded. If the packet does not arrive within the maximal delay, the gateway will deliver the duplicate ACK.

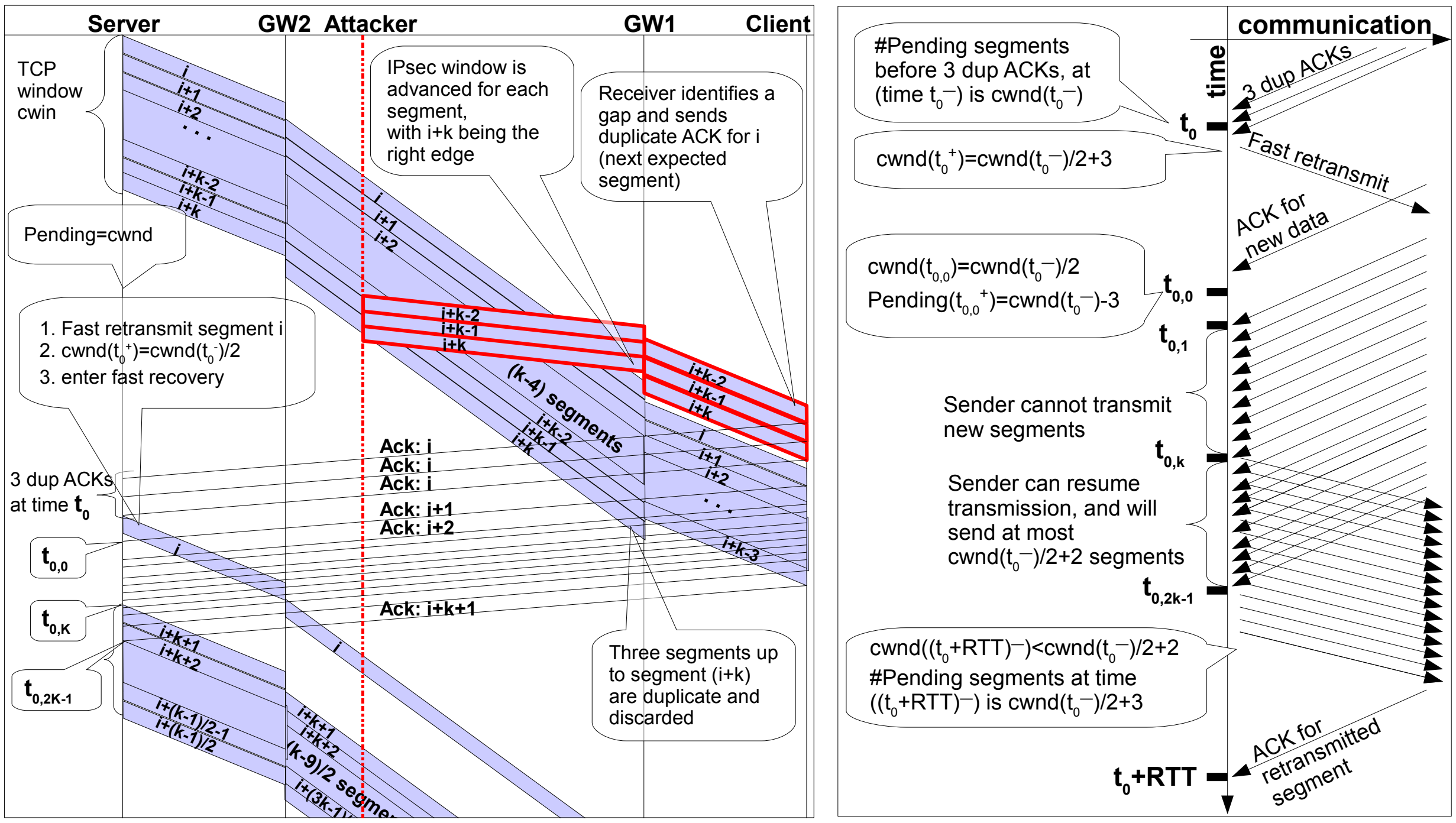

(a) TCP sender transmits a window of $k+1$ segments, and the attacker speeds up last three segments, i.e., segments with sequence numbers $i+k-2, i+k-1, i+k$, such that they arrive before $1^{st}$ segment in the window, i.e., segment with sequence number $i$. These segments are passed by the gateway to the client (since they are authentic and not a replay), and trigger three duplicate ACKs for $i^{th}$ segment at the receiver. Upon receipt of three duplicate ACKs the sender retransmits 'lost' segment $i$, and reduces its transmission rate.

(b) At time $t_0$ the sender receives three duplicate ACKs (this initiates first attack epoch); also note that $cwnd = pending$. The sender fast retransmits lost segment, and reduces its transmission rate. The sender cannot resume transmission of further segments since the number of pending segments is larger than the congestion window size. After receipt of sufficient number of ACKs the sender can resume transmission of new segments.

**Fig. 7.** Throughput degradation attack on communication over TCP even for an infinitely large IPsec anti-replay window, and packets are not discarded by IPsec implementation at the receiving gateway. Attacker duplicates and speeds up three segments, resulting in gap in sequence numbers at the receiver, which responds with duplicate ACKs. Upon receipt of three duplicate ACKs, the sender reduces its sending rate.

The receiving gateway will store a typical delay[7] from each VPN to receiver, and for each incoming segment from the Internet the receiving gateway will check if delay for the arriving segment is less than the typical delay on network between gateways. If yes, the gateway will record receipt time, and will set an alert flag on. Otherwise it will proceed as usual. Then if a duplicate ACK has arrived, the gateway will check if alert flag is turned on, and will store the duplicate ACK and each subsequent duplicate ACK with the same sequence number. If within the typical delay from sender the segment with that sequence number is received, the gateway will discard duplicate ACK(s). Otherwise it will forward the duplicate ACKs. The detailed implementation of the reordering tolerant tunneling protocol (at the receiver, i.e., gateway $GW1$ in Figure 1(a)) is in Algorithm 1. The receiving gateway has to procedures, one receiving packets destined from the local network to the Internet (Outgoing Packet), and another from the Internet to the local network

[7] The gateway will estimate the typical delay between itself and the sending gateway, following the existing approaches that estimate TCP timeout.

(Incoming Packet). The sending gateway ($GW2$ in Figure 1(a)) implementation remains the same and only adds a timestamp on each outgoing packet and authenticates it.

---

**Algorithm 1:** Implementation of reordering tolerant tunneling mechanism in the receiving gateway.

```
Input: pkt
Data: typicalDelay = setTypicalDelay(); alert = 0

Incoming Packet
    foreach pkt do
        if dupACKcnt ≠ 0 ∧ pkt.SN == ackSN then
            alert = 0;
            free pending ACKs;
            stop timer;
        end
        else
            if ((pkt.delay < typicalDelay) ∧
            ((expected sequence number) < pkt.SN))
            then
                pktSN ← pkt.SN;
                rcptTime ← pkt.time;
                alert = 1;
            end
        end
    end
end

Outgoing Packet
    dupACKcnt = 0;
    ackSN = 0;
    if (alert = 1) ∧ (timer + rcptTime = typicalDelay)
    then
        send all pending duplicate ACKs;
    end
    foreach ack do
        if ack.SN = ackSN then
            dupACKcnt + +;
            set timer;
            if (ack.data) then
                send data with lower SN
            end
        end
        else
            ackSN = ack.SN;
            if dupACKcnt ≠ 0 then
                dupACKcnt = 0;
            end
        end
    end
end
```

---

The reordering tolerant mechanism in firewall (in Algorithm 1) ensures optimal performance for communication over TCP when reordering occurs due to benign network conditions, as well as as a result of attack. If the firewall receives a segment with a higher sequence number (SN) than the one it expects to receive, it will turn the alert flag on. Once duplicate ACK arrives, it is stored and not forwarded. When the legitimate packet arrives, all pending duplicate ACKs are freed and alert is turned off. If legitimate packet does not arrive after a period of time, (which is a function of the speed-up of the adversary w.r.t. delay on network-i.e., the packet was probably lost) send the delayed duplicate ACKs[8]

[8] Since TCP is full duplex, ACKs may contain data. In case a duplicate ACK contains data, we store the duplicate ACK, and create a new packet to transfer the data. The packet will have a sequence number (SN) of a previously sent sequence number. TCP implementations ignore outdated sequence numbers.